\documentclass[amssymb,amsmath,prd,twocolumn,superscriptaddress,nofootinbib]{revtex4-1}

\usepackage{graphicx}
\usepackage{bm}
\usepackage{amssymb,amsmath}
\usepackage{mathrsfs}
\usepackage{latexsym}
\usepackage{color}
\usepackage[normalem]{ulem}
\usepackage{dcolumn}
\usepackage[colorlinks=true,citecolor=blue,urlcolor=blue]{hyperref}
\usepackage[usenames,dvipsnames]{xcolor}
\usepackage{xspace}
\usepackage{graphicx}

\allowdisplaybreaks

\newcommand{\CIT}{\affiliation{Department of Physics, California Institute of Technology, Pasadena, California 91125, USA}}
\newcommand{\CITLab}{\affiliation{LIGO Laboratory, California Institute of Technology, Pasadena, California 91125, USA}}

\definecolor{kcmagenta}{rgb}{0.54, 0.17, 0.88}

\begin{document}

\title{Uncertainty limits on neutron star radius measurements with gravitational waves}

\author{Katerina Chatziioannou} \CIT \CITLab

\date{\today}

\begin{abstract}
Upcoming observing campaigns with improved detectors will yield numerous detections of gravitational waves from neutron star binary inspirals. 
Rare loud signals together with numerous signals of moderate strength promise stringent constraints on the properties of neutron star matter, with a projected radius 
statistical uncertainty of $50-200$m with ${\cal{O}}(2000)$ sources. 
Given this precision we revisit all analysis assumptions and identify sources of systematic errors, quantify their impact on radius extraction, and discuss their relative importance and ways to mitigate them. 
\end{abstract}

\maketitle

\section{Introduction}
\label{sec:intro}

Astronomical observations constrain the macroscopic properties of neutron stars (NSs) 
and the behavior of dense, cold matter~\cite{Lattimer:2015nhk,Oertel:2016bki,Baym:2017whm}. 
Among them, gravitational wave (GW) observations of binary NS (BNS) inspirals lead to a measurement of the masses and tidal properties of the stars~\cite{Chatziioannou:2020pqz,Dietrich:2020eud}.
The binary masses affect the evolution of the early, minute-long signal, while each NS's response to its companion's gravitational tidal field leaves an imprint on the signal during the final coalescence stages. 

The mutual tidal interaction induces a quadrupole moment on each NS which accelerates the GW inspiral. The leading-order effect
 is quantified through the tidal deformability $\Lambda \equiv (2/3) k_2 (R/m)^{5}$~\cite{Flanagan:2007ix,Hinderer:2009ca}, 
 where $k_2$ is the Love number of a NS with mass $m$ and radius $R$. 
 Measuring $\Lambda$ offers complementary information to traditional radius measurements through $k_2$, 
 while the strong mass dependence implies tighter constraints for less massive, and thus more deformable, NSs.
The first BNS coalescence detected with GWs, GW170817~\cite{TheLIGOScientific:2017qsa}, had a signal-to-noise ratio (SNR)
of $\rho=32$, and a combined tidal deformability 
$\tilde{\Lambda}$~\cite{Favata:2013rwa,Wade:2014vqa} of $300^{+420}_{-230}$ at the 90\% credible level~\cite{Abbott:2018exr,Abbott:2018wiz}. This result has been extensively
shown to disfavor very stiff nuclear matter and large NS radii and tidal deformabilities~\cite{Annala:2017llu,Raithel:2018ncd,Radice:2018ozg,Coughlin:2018fis,Essick:2019ldf,Landry:2018prl,PhysRevLett.121.091102,Raaijmakers:2019dks,Capano:2019eae,Dietrich:2020efo,Landry:2020vaw,Essick:2020flb,LIGOScientific:2019eut}
 and to be in agreement with further astronomical and terrestrial constraints~\cite{Landry:2020vaw,Legred:2021hdx,LIGOScientific:2019eut,Raaijmakers:2021uju,Essick:2020flb,Essick:2021kjb,Biswas:2021yge,Biswas:2020puz,Miller:2021qha,Al-Mamun:2020vzu,Reed:2021nqk,Tews:2019ioa,Tews:2019cap}.
 
Scheduled or planned GW detector upgrades and observing campaigns~\cite{Aasi:2013wya} are expected to yield further BNS detections and improve on the overall constraints
by both combining information from multiple events of different masses~\cite{Lackey:2014fwa,Abdelsalhin:2017cih,HernandezVivanco:2019vvk,Landry:2020vaw} and detecting louder
signals. The uncertainty in tidal parameters scales as $\rho^{-1}$~\cite{Wade:2014vqa}, however, the tidal deformability affects the 
GW signal primarily during the late stages of the coalescence for 
frequencies $\gtrsim 400$Hz. Lower frequencies are essential for identifying the signal and estimating the masses, but tidal inference primarily relies on the 
SNR accumulated, and thus the expected detector performance, at this frequency range.

Since the SNR is inversely proportional to the detector sensitivity, we can make rough estimates about the expected constraints on the NS tidal deformability. 
The LIGO-Livingston~\cite{TheLIGOScientific:2014jea} strain sensitivity in the relevant frequencies improved by a factor of $\sim$2 between the
 second~\cite{GW170817Samples,O2catalogPErelease} and third observing runs~\cite{Abbott:2020uma,LIGOScientific:2020ibl}. Design sensitivity could bring another factor of $\sim$1.5~\cite{aLIGO_design_updated}, with the A+ and Voyager~\cite{LIGO:2020xsf,ASD_curves18} upgrades yielding improvements by factors of $\sim$2 and $\sim$1.5  respectively. Next generation 3G detectors are envisioned to have $\sim$10 times better strain sensitivity than advanced LIGO~\cite{Evans:2016mbw,Reitze:2019dyk,Reitze:2019iox,2011CQGra..28i4013H,2010CQGra..27h4007P}. Though these improvements are not uniform across the frequency band, they roughly suggest that a GW170817-like event would have an SNR ($\tilde{\Lambda}$ 90\% uncertainty) of $100 (200)$ at design sensitivity, $200 (100)$ with A+, $300 (66)$ with Voyager, and $1000 (20)$ with 3G detectors. 

The total number of sources observed per detector upgrade depends on the BNS merger rate and its distribution with redshift with a current estimate of $320^{+490}_{-240}  \mathrm{Gpc}^{-3} \mathrm{yr}^{-1}$ mergers uniformly distributed in comoving volume and for uniform NS masses in $(1,2.5)M_{\odot}$~\cite{LIGOScientific:2020kqk}. 
With these specifications, the median merger rate, and a detection SNR threshold of $12$ we use the {\tt GW-Toolbox}~\cite{Yi:2021wqf} 
to estimate ${\cal{O}}(10)$ detections per year at design sensitivity, ${\cal{O}}(100)$ for A+, and ${\cal{O}}(500)$ for Voyager. The latter is $7$ times lower than the  
estimate of~\cite{LIGO:2020xsf}, likely due to differences in the merger rate and SNR threshold. 
Predictions for 3G detectors are subject to the uncertain redshift distribution of mergers, but estimates suggest ${\cal{O}}(1000)$ sources per day~\cite{Regimbau:2012ir,Sachdev:2020bkk}. 

We estimate the total SNR accumulated by such a catalog of BNS detections by adopting a $\rho^{-4}$ distribution~\cite{Chen:2014yla}, appropriate for detectors up to Voyager that get the majority of their sources from redshifts up to $0.3$~\cite{LIGO:2020xsf}.
We simulate $100$ source catalogs and compute the total SNR by summing the per-event SNRs in quadrature to find a median total SNR of 60(200)[450]\{650\} with 10(100)[500]\{1000\} sources.
The total SNR from 3G detectors is expected to be ${\cal{O}}(10^4)$~\cite{Haster:2020sdh}. These estimates are conservative if dedicated high-frequency
detectors join the network~\cite{Bailes:2019oma,2021PhRvD.103b2002G}.

\begin{figure}
\centering
\includegraphics[width=0.49\textwidth]{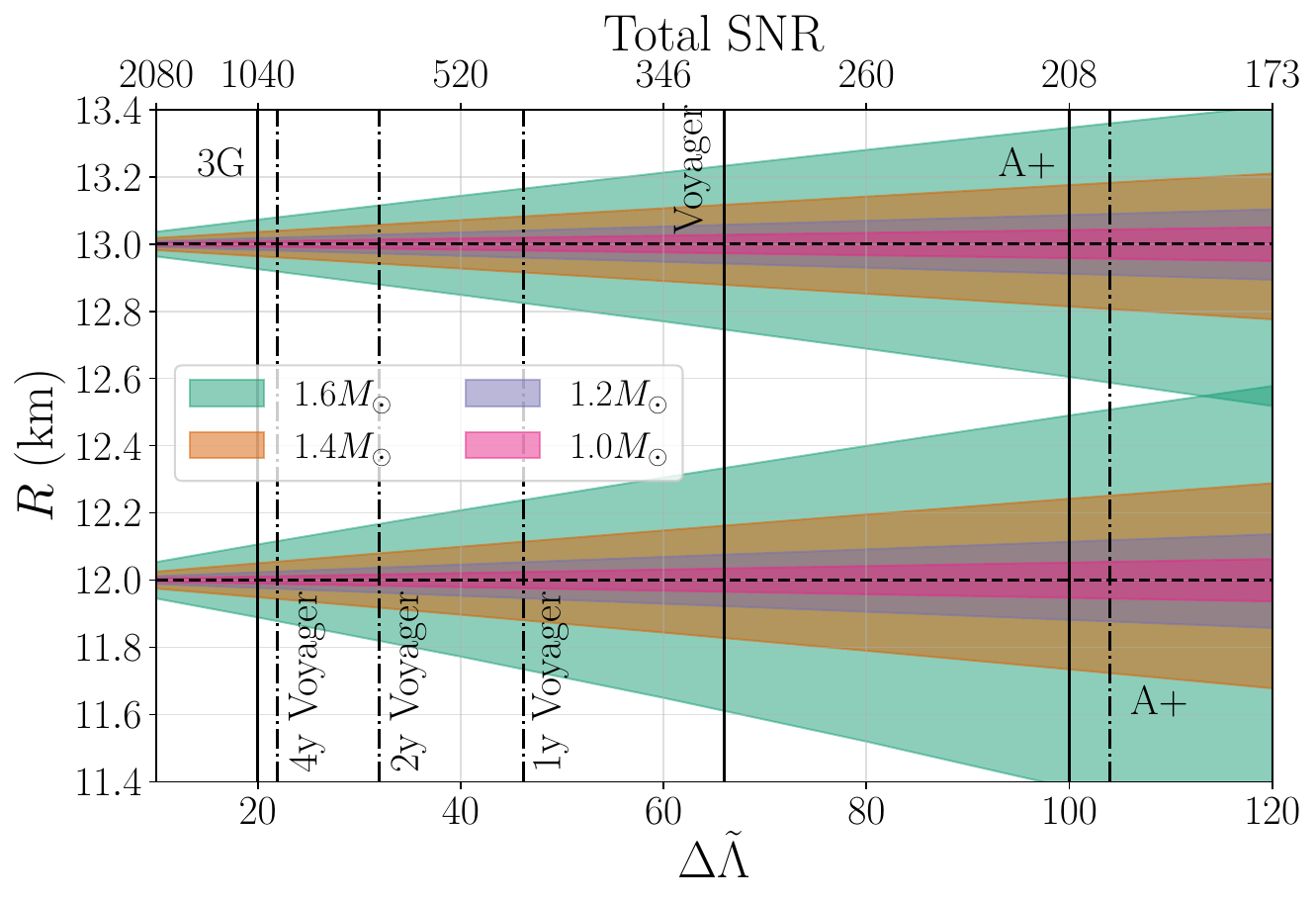}
\caption{Radius 90\% uncertainty as a function of the uncertainty in the combined tidal deformability for different NS masses and radii. The top axis gives the corresponding SNR, achieved with either a single source or a combination of sources. Estimates are based on GW170817's $\Delta\tilde{\Lambda}\sim650$ at SNR $32$. Solid (dot-dashed) vertical lines are projected results from a GW170817-like event (a catalog of BNSs) with different detectors and observing durations.}
\label{fig:deltaR}
\end{figure}

The corresponding NS radius accuracy is shown in Fig.~\ref{fig:deltaR} for different NS masses and radii, assuming we can perfectly convert a ($m,\tilde{\Lambda}$) measurement to $R$. Since we are interested in the expected radius uncertainty and not the details of the calculation, we use the relations of~\cite{Maselli:2013mva,Yagi:2015pkc,Chatziioannou:2018vzf,Carson:2019rjx} between the masses, tides, and radius. In practice such relations carry additional systematic uncertainty, so approaches that model the whole NS equation of state hierarchically would be preferred in the regime of informative measurements. However, our goal here is an estimate of the expected radius uncertainty and not the details of which analysis can achieve it, so such relations are appropriate.

The $\Lambda \sim (R/m)^{5}$ scaling results in increased radius uncertainty with higher/lower NS mass/radius due to intrinsically weaker binary tidal interactions. An SNR of $1000$, achieved either by observing a GW170817-like event with 3G detectors
or with $4$ years of Voyager operation, would result in a measurement of $\tilde{\Lambda}$ to $20$ and $R$ to $50- 200$m for different NS masses.
A more moderate total SNR of $200$ from $\sim$100 sources would lead to a $\tilde{\Lambda}$ ($R$) uncertainty of $100$ ($1$km) at
$1.6M_{\odot}$, consistent with the more detailed simulations of~\cite{Landry:2020vaw}. The final constraint achieved on NS radii will be a combination of these per-mass estimates depending on the astrophysical NS mass distribution.

The above are not detailed predictions about the expected NS constraints from future detectors; such estimates would require a precise treatment of --among others-- the
BNS merger rate, its redshift distribution, the NS mass distribution, the broadband detector performance, the network duty cycle, etc. 
However, they provide
a projection that GWs could result in a $\sim$100m radius measurement within the decade with Voyager and beyond with 3G detectors. 
This radius constraint would also improve if effects such as dynamical tides~\cite{Pratten:2019sed,Schmidt:2019wrl,Ma:2020oni} are detected as they are qualitatively different than the standard adiabatic tides considered here and not captured by the $\rho^{-1}$ scaling.
Reaching this projected precision relies on ascertaining that every aspect of the GW analysis induces potential systematic errors that are fully quantified and brought below 
statistical uncertainties.  

\section{Gravitational wave analysis}
\label{sec:analysis}

Analysis of GW data $d$ to extract source parameters $\theta$ relies on modeling the signal with a waveform template $h(\theta)$ under some model for the detector noise. The likelihood function in the frequency domain is~\cite{Veitch:2014wba,Romano:2016dpx}
\begin{equation}
\mathrm{log}L\sim -\frac{1}{2}(d-h(\theta)|d-h(\theta)),
\end{equation}
with the noise-weighted inner product
\begin{equation}
	(a|b) \equiv 2 \int \frac{a^*(f) b(f) + b^*(f) a(f)}{S_{n}(f)} df, \label{eq-innerprod}
\end{equation}
where an asterisk denotes complex conjugation and $S_{n}(f)$ is the power spectral density (PSD) of the noise. The likelihood and a prior for $\theta$ give the posterior probability.

The above allows us to identify the ingredients of parameter estimation: 
\begin{itemize}
\item[(i)] the data $d$, 
\item[(ii)] the noise PSD $S_n(f)$, 
\item[(iii)] the waveform model $h(\theta)$, 
\end{itemize}
as well as the main assumptions:
\begin{itemize}
\item[(iv)] the detector noise is stationary, leading to a diagonal noise covariance matrix and an inner product that is a one-dimensional frequency integral, and
\item[(v)] the detector noise is gaussian, which leads to the Gaussian functional form of the likelihood.
\end{itemize}
Each of the above introduces systematic uncertainties that will affect inference at some level.

\section{Assumption: Gaussian noise}
\label{sec:gaussian}

The functional form of the likelihood is dictated by the assumption of gaussian detector noise. Gaussianity can be violated by  instrumental artifacts, known as glitches, or multiple GW signals temporally overlapping. Glitches are a common occurrence, with a rate of $\lesssim 1$ per minute in the LIGO detectors in O3a~\cite{LIGOScientific:2020ibl} and already coinciding with signals, notably GW170817~\cite{TheLIGOScientific:2017qsa,Pankow:2018qpo}. Given the typical duration of BNS signals in the LIGO band of a few minutes and the rate of glitches, we expect such BNS-glitch overlaps to be quite common. Overlapping signals are expected to be rare in advanced LIGO but a possibility for 3G detectors~\cite{Pizzati:2021gzd,Samajdar:2021egv,Relton:2021cax,Antonelli:2021vwg}.

The temporal coincidence of glitches and signals has led to the development of mitigation techniques that simultaneously model the signal and the glitch~\cite{Chatziioannou:2021ezd} or use auxiliary channel information~\cite{Tiwari:2015ofa,Meadors:2013lja,Ormiston:2020ele,Mogushi:2021deu}. 
In the context of tidal inference, glitches are relevant when overlapping with the signal at frequencies $\gtrsim$400Hz.
Though O3a was dominated by glitches with peak frequencies below 100Hz~\cite{LIGOScientific:2020ibl}, those glitches extend to higher frequencies and improved detector sensitivity could bring 
new glitch families. A prominent glitch will be modeled together with the signal~\cite{Chatziioannou:2021ezd}, leading to unbiased signal parameters. However, this does not preclude the possibility of a stealth bias~\cite{Vallisneri:2013rc}, where the glitch is not loud enough to be identified but could still affect tidal inference.

\begin{figure}
\centering
\includegraphics[width=0.49\textwidth]{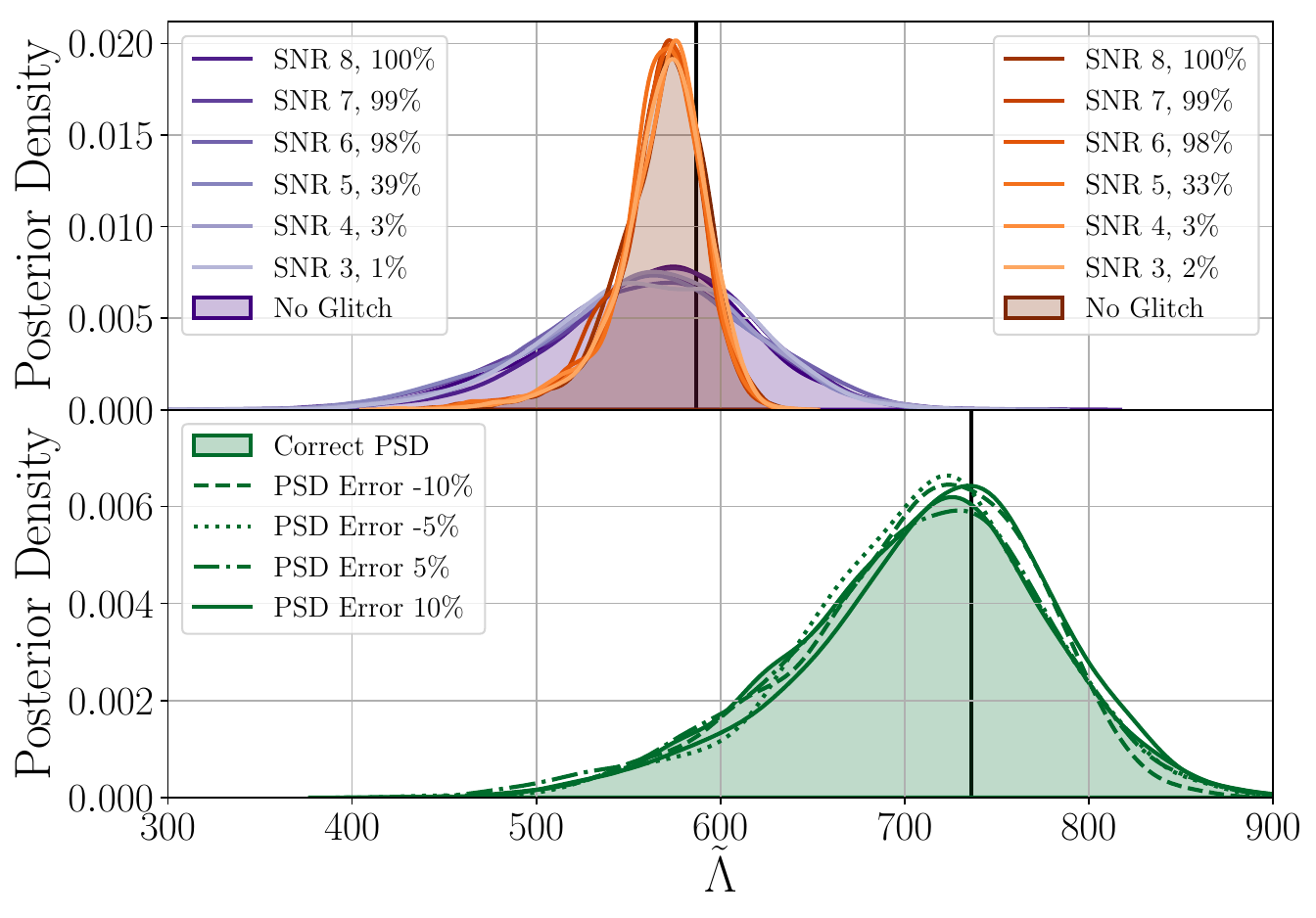}
\caption{Marginalized posterior for $\tilde{\Lambda}$ examining the effect of non-Gaussian residuals (top) and misestimated noise PSD (bottom). Black vertical lines give the injected values. Purple (orange) shade distributions correspond to simulated signals of SNR 140 (420) in data with glitches. Green shade distributions correspond to signals analyzed with different misestimated high-frequency PSDs. Legends give the glitch SNR and the identification probability, defined as the percentage of posterior samples that model the glitch (top), and the level of PSD misestimation (bottom). }
\label{fig:LambdaT}
\end{figure}

We explore this possibility by simulating BNS signals with SNR 140 and 420 in a zero-noise realization. We break the Gaussianity assumption by adding a glitch in the LIGO-Livingston data given by a sine-gaussian that overlaps with the BNS signal at 700Hz, the most interesting frequency region for tidal inference. The quality factor of the glitch is 20, resulting in an instrumental transient that resembles the underlying signal but is present in only one detector. We vary the SNR of the glitch and simultaneously model the signal with waveform templates and the glitch with wavelets~\cite{Chatziioannou:2021ezd}. 

Figure~\ref{fig:LambdaT} shows the resulting $\tilde{\Lambda}$ posterior also compared to the case of no glitch and purely Gaussian noise. Uncertainties are  consistent with the signal SNR and the projections of Fig.~\ref{fig:deltaR}.
In all cases the correct value is recovered regardless of the SNR of the glitch. The skewed shape of the posterior is due to a 
 correlation between $\tilde{\Lambda}$ and the binary mass ratio causing the mean of the posterior to be to the left of the injected value; we have verified that the two-dimensional posterior peaks at the injected value.
For glitches with $\rho\gtrsim$6 the probability of glitch identification is $\sim$1. For lower glitch SNR the probability drops, with only $3\%$ of the posterior samples identifying a glitch of $\rho$=4. Despite this, the $\tilde{\Lambda}$ posterior remains unbiased, suggesting no stealth bias: if the glitch is too quiet to be identified, it is also too quiet to bias tidal inference. This example considers one glitch and signal and it is possible that different glitch morphologies
could prove more problematic. However, it shows that glitch mitigation techniques are already in place and able to handle instrumental artifacts in the data. 

The other possibility is that of a secondary signal overlapping with the primary BNS. 
The most problematic scenario occurs for signals separated by less than $0.1$s~\cite{Pizzati:2021gzd,Samajdar:2021egv,Relton:2021cax,Antonelli:2021vwg} hence the secondary signal could overlap with the tidally-affected region
of the primary signal. Biases could occur also if the secondary signal is subthreshold and not individually resolvable~\cite{Relton:2021cax}. This case requires a joint analysis of the two signals, similar to the joint signal and glitch analysis from above, and also similar to techniques developed
for space-based detectors whose data contain millions of overlapping signals~\cite{Cornish:2005qw,Littenberg:2011zg,Littenberg:2020bxy}.

In summary, non-Gaussian features in the data are unavoidable, but they can be modeled simultaneously with the primary signal to ensure unbiased parameter recovery. What is more, any systematic bias from non-Gaussian data will be unique to each BNS. The bias will therefore not accumulate when multiple events are combined, though it might 
affect individual very loud sources.

\section{Assumption and Ingredient: Stationary noise with a known PSD}
\label{sec:noise}

The inner product of Eq.~\eqref{eq-innerprod} is based on the assumption that the Gaussian detector noise is stationary: its mean and covariance are constant in time
 for the duration of the analyzed data segment. 
The frequency-domain covariance matrix is then diagonal with a (assumed known) variance related to $S_n(f)$. Neither noise stationarity nor a perfect knowledge of $S_n(f)$ are
strictly true, introducing a potential systematic in parameter estimation.

Tidal effects influence the signal for a few tens of milliseconds, during which the noise is most likely stationary. However, the entire signal lasts for longer as a GW170817-like 
signal is 2(5)\{10\} minutes from merger at 23(16)\{12\}Hz, putting a strain on the stationarity assumption. Efforts to subtract
nonstationary noise~\cite{Vajente:2019ycy,Yu:2021vvm} or correct for it at the analysis level~\cite{Zackay:2019kkv} are under way,
with a further option of abandoning the frequency domain altogether~\cite{Isi:2021iql,Cornish:2020odn}.
Spectral lines in the data~\cite{Littenberg:2014oda,LIGOScientific:2019hgc,Huang:2020ysn} and finite analysis segments~\cite{Talbot:2021igi} can also introduce nondiagonal terms in the covariance matrix.

Under the assumption of stationarity the noise PSD itself can be computed using either off-source~\cite{Veitch:2014wba,Romero-Shaw:2020owr} 
data that assume stationarity from even longer data segments or it can be modeled based
on on-source data~\cite{Littenberg:2014oda,Chatziioannou:2019}. Uncertainty in the PSD estimation can also be marginalized over~\cite{Rover:2008yp,Rover:2011qd,Talbot:2020auc,Chatziioannou:2021ezd}. PSD errors could cause parameter biases~\cite{Chatziioannou:2019}, however these should primarily affect the amplitude of the GW signal and less so its phase evolution. 

To test this we simulate a BNS in Gaussian noise and analyze it with a misestimated
noise PSD in the $(400,1000)$Hz range. The PSD is based on the LIGO design sensitivity and we alter its strain sensitivity in the relevant frequency range by
some percentage compared to the true value used for the simulated data.
The resulting $\tilde{\Lambda}$ posterior is shown in Fig.~\ref{fig:LambdaT} showing PSD misestimation does not affect tidal 
parameter recovery for PSD relative errors of up to $\pm 10\%$.

\section{Ingredient: Data}
\label{sec:data}

The data $d$ correspond to the relative displacement of the interferometer test masses, tracked through interfering laser light incident on photodetectors.
Converting the photodetector output to strain is achieved through a calibration process whose
uncertainties could affect parameter estimation if left unaccounted for~\cite{Lindblom:2009un,Vitale:2011wu}.
Detector calibration relies on detector strain induced by photon calibrators~\cite{Karki:2016pht}, resulting in an estimate
for the systematic error and corresponding statistical uncertainty for the detector frequency-domain amplitude and phase response~\cite{LIGOScientific:2016xax,2017PhRvD..96j2001C,Sun:2020wke,Sun:2021qcg,VIRGO:2021umk}. 

During first half of O3 the calibration uncertainty (systematic and statistical) was determined to be no more than 4$^\circ$ in phase in the LIGO detectors at the 68\% level, corresponding to 7$^\circ$ at the 90\% level~\cite{Sun:2020wke}, with similar estimates for the second half of O3~\cite{Sun:2021qcg}. A conservative phase calibration error of 10$^\circ$ is compared in Fig.~\ref{fig:DeltaPhi} against the GW phase shift for different NS radii. The phase calibration uncertainty is comparable to the GW dephasing induced by a $100-200$m change in the radius.

Astrophysical parameter estimation studies marginalize
over calibration uncertainty~\cite{LIGOScientific:2016vlm}, a procedure that effectively increases measurement uncertainties, though the result is small at current sensitivities. Though
calibration uncertainty is typically treated as being uncorrelated between different frequencies, using a physical calibration model that correctly encodes calibration error across frequencies could further mitigate the effect on parameter estimation~\cite{Vitale:2020gvb,Payne:2020myg}. Improvements in photon calibration~\cite{Bhattacharjee:2020yxe}, alternative methods such as the Newtonian calibrator~\cite{Estevez:2018zdr,Ross:2021apo}, and even astrophysical calibration~\cite{Pitkin:2015kgm,Essick:2019dow} could reduce the impact of calibration error which is currently comparable to the target radius uncertainty of $100$m.

\section{Ingredient: Waveform model}
\label{sec:model}

The final ingredient of the analysis, and the most commonly considered one in the context of systematics, is the waveform model~\cite{Dietrich:2020eud}. The effect of waveform systematics has been investigated for GW170817 by employing a diverse set of waveform models including post-Newtonian~\cite{Flanagan:2007ix,Vines:2011ud}, effective-one-body~\cite{Damour:2009vw,Bernuzzi:2014owa,Nagar:2018zoe,Gamba:2020ljo,Hinderer:2016eia,Steinhoff:2016rfi}, and phenomenological models calibrated to numerical relativity simulations~\cite{Dietrich:2017aum,2019PhRvD..99b4029D,Dietrich:2019kaq,Kawaguchi:2018gvj}. The main conclusion is that current statistical uncertainties dominate over
waveform systematics~\cite{Abbott:2018wiz,LIGOScientific:2018mvr}.

Studies of simulated signals using a wide variety of waveforms suggest, however, that waveform systematics could become significant for 
 $\rho \gtrsim$100~\cite{Dietrich:2020eud,Dudi:2018jzn,Samajdar:2019ulq,Gamba:2020wgg}, corresponding to GW170817 at design sensitivity or $50$ sources at the A+ timescale.
 Waveform biases increase with $\tilde{\Lambda}$, and thus less massive or bigger NSs, and could be due to modeling error in the point-particle
 or the tidal sectors of the waveform~\cite{Dietrich:2020eud}.  Additionally, numerical errors in the numerical relativity simulations used to calibrate the waveform models could influence results at a similar SNR~\cite{Dudi:2018jzn} as in some cases waveform and simulation errors can be comparable~\cite{Bernuzzi:2011aq,Bernuzzi:2016pie,Kiuchi:2017pte,Dietrich:2018upm,Dietrich:2017feu,Dietrich:2018phi}.
 This level of systematic error will be comparable to statistical errors for detectors with A+ sensitivity, once the statistical uncertainty in $\tilde{\Lambda}$ ($R$) reaches 200 (0.5-1km) at the 90\% credible level.

\begin{figure}
\centering
\includegraphics[width=0.49\textwidth]{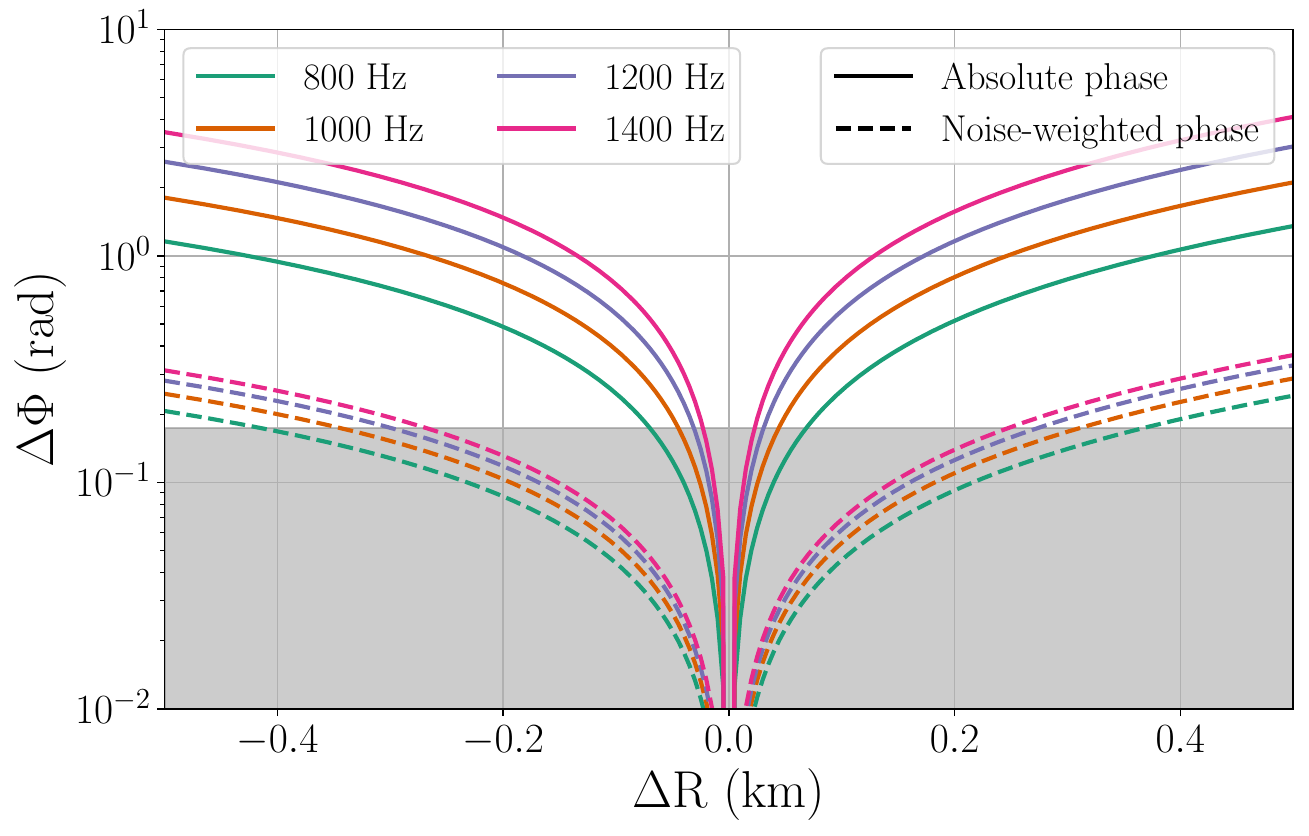}
\caption{GW frequency-domain phase difference between signals from BNSs with different radii relative to $13$km up to different frequencies. Solid (dashed) lines denote absolute (noise-weighted) phase differences. The shaded region marks a $10^{\circ}$ calibration uncertainty, relevant only for the absolute phase difference. Unbiased tidal and radius inference hinges on waveform models achieving phase evolution accuracies better than the induced tidal effect.}
\label{fig:DeltaPhi}
\end{figure}

We quantify the waveform accuracy required to achieve the radius measurement projected for Voyager and 3G detectors in Fig.~\ref{fig:DeltaPhi}. Using the {\tt IMRPhenomD\_NRTidalv2}~\cite{Dietrich:2019kaq} waveform model we compute the frequency-domain phase difference between signals emitted by BNSs of $1.4M_{\odot}$ as a function of the NS radius
relative to $13$km. A $0.5$km radius difference induces a $3-4$rad dephasing; waveform systematics need to be kept below this level to achieve such radius accuracy. Errors in numerical simulations are typically quoted at $\sim$1rad~\cite{Dietrich:2017feu}, though these refer to time-domain phase and are not directly comparable.

The dephasing is larger at higher frequencies where the detector sensitivity decreases, thus making absolute phase differences less informative.
As an alternative we also plot a noise-weighted phase difference, derived in~\cite{Sampson:2014qqa} as the effective cycles of phase that a specific effect (here the tidal deformation) contributes to the waveform. The effective cycles of phase
are related to an upper limit on the Bayes factor that the effect in question is detectable in the data~\cite{Sampson:2014qqa}. We use the LIGO design sensitivity, though the level of the PSD does not affect the noise-weighted phase difference, only its shape.

The noise-weighted phase difference is a less sensitive function of the upper frequency cutoff, but it still increases for frequencies up to $1400$Hz. A $0.5$km change in the radius leads to a noise-weighted phase change of  $\sim$0.3rad, again setting the threshold for waveform accuracy in other to achieve such a radius measurement.

\section{Conclusions}
\label{sec:conclusions}

The astounding precision expected to be achieved with GW observations in the next decade(s)~\cite{Adhikari:2019zpy} brings to the forefront all analysis assumptions and ingredients to be examined under the light of systematic errors. In the context of BNSs we study the frontier for measuring NS properties by quantifying the limiting effect of various
sources of systematics. Improved waveform models are needed for a radius measurement of $\Delta R\sim$1km at the timescale of A+, while improvements in detector calibration are needed at the timescale of Voyager for $\Delta R\sim$200m. Both waveform and calibration errors would introduce correlated biases for different sources, and thus increasingly affect combined constraints. 

The importance of non-Gaussian noise depends on the type of glitches in future detectors, but methods to simultaneously model signals and glitches or multiple
signals will mitigate the effect. The effect of nonstationary noise depends on the detector low frequency performance and hence the signal length. Misestimating the noise PSD has a negligible effect on tidal inference. Non-Gaussianity, nonstationarity, and PSD misestimation would likely affect each signal in a unique way, and thus not accumulate in combined constraints.

This discussion concerns extracting binary parameters, namely masses and tidal deformability, from GW data. The projected constraints were translated to 
NS radius assuming a perfect conversion method for illustration. 
In reality, additional analysis steps are required to combine information from multiple sources and obtain constraints on NS radii or features such as 
phase transitions~\cite{DelPozzo:2013ala,Agathos:2015uaa,Chatziioannou:2015uea,Han:2018mtj,Chatziioannou:2019yko,Han:2020adu,Zhang:2019fog,Pang:2020ilf,Drischler:2020fvz}.
This is achieved through hierarchical inference~\cite{Mandel:2009pc,Mandel:2018mve} which accounts for
parameters correlations such as the ones between $\tilde{\Lambda}$ and mass ratio or the distance and the inclination, statistical uncertainties, and selection effects. Additional systematic errors here would be the NS equation of state model employed~\cite{Legred:2021hdx,Greif:2018njt}, the estimation of the selection function~\cite{LIGOScientific:2018jsj,LIGOScientific:2020kqk,Wysocki:2018mpo,2019RNAAS...3...66F}, and the population model for NS spins and masses~\cite{Chatziioannou:2020msi,Landry:2021hvl}. 
The latter, if neglected, could bias the NS equation of state after 25-50 events~\cite{Wysocki:2020myz} but the effect can be fully addressed by simultaneously inferring the NS mass and spin distribution with the equation of state~\cite{Wysocki:2020myz,Golomb:2021tll}.

Finally, this study examines only GW data and their analysis. X-ray observations~\cite{Miller:2019cac,Riley:2019yda,Miller:2021qha,Riley:2021pdl,Bogdanov:2019ixe,Bogdanov:2019qjb,Bogdanov:2021yip} 
will provide additional information about NS properties yielding overall tighter constraints. 
The relation between systematic errors in the different observations and their effect on the decreasing statistical uncertainty needs to be explored.

\section{Acknowledgements}
\label{sec:acks}

We thank the participants of the Aspen Center program ``Exploring Extreme Matter in the Era of Multimessenger Astronomy: from the Cosmos to Quarks" for numerous discussions that motivated this study. We thank Max Isi for comments on the manuscript.
This work was initiated at Aspen Center for Physics, which is supported by National Science Foundation grant PHY-1607611.
This work was partially supported by a grant from the Simons Foundation.
This research has made use of data, software and/or web tools obtained from the Gravitational Wave Open Science Center (https://www.gw-openscience.org), a service of LIGO Laboratory, the LIGO Scientific Collaboration and the Virgo Collaboration.
Virgo is funded by the French Centre National de Recherche Scientifique (CNRS), the Italian Istituto Nazionale della Fisica Nucleare (INFN) and the Dutch Nikhef, with contributions by Polish and Hungarian institutes.
This material is based upon work supported by NSF's LIGO Laboratory which is a major facility fully funded by the National Science Foundation.
The authors are grateful for computational resources provided by the LIGO Laboratory and supported by National Science Foundation Grants PHY-0757058 and PHY-0823459.
Software: {\tt gwpy}~\cite{duncan_macleod_2020_3598469}, {\tt matplotlib}~\cite{Hunter:2007}, {\tt gwtoolbox}~\cite{Yi:2021wqf}.

\bibliography{OurRefs}

\end{document}